\documentclass[aps,prl,reprint,groupedaddress,longbibliography]{revtex4-2}
\usepackage{amsmath,amssymb,epsfig,bbm,bm,verbatim,MnSymbol,mathtools,amsthm}
\usepackage[colorlinks=true,citecolor=blue]{hyperref}

\begin{document}


\title{Geometric Bounds on the Power of Adiabatic Thermal Machines}


\author{Joshua Eglinton}
\author{Kay Brandner}
\affiliation{
School of Physics and Astronomy,
University of Nottingham,
Nottingham NG7 2RD,
United Kingdom\\
Centre for the Mathematics and Theoretical Physics of Quantum Non-equilibrium Systems,
University of Nottingham,
Nottingham NG7 2RD, 
United Kingdom
}


\date{\today}

\begin{abstract}
We analyze the performance of slowly driven meso- and micro-scale refrigerators and heat engines that operate between two thermal baths with small temperature difference. 
Using a general scaling argument, we show that such devices can work arbitrarily close to their Carnot limit only if heat-leaks between the baths are fully suppressed. 
Their power output is then subject to a universal geometric bound that decays quadratically to zero at the Carnot limit. 
This bound can be asymptotically saturated in the quasi-static limit if the driving protocols are suitably optimized and the temperature difference between the baths goes to zero with the driving frequency. 
These results hold under generic conditions for any thermodynamically consistent dynamics admitting a well-defined adiabatic-response regime and a generalized Onsager symmetry. 
For illustration, we work out models of a qubit-refrigerator and a coherent charge pump operating as a cooling device. 
\end{abstract}


\maketitle

Dimensionless figures of merit, such as the efficiency of a heat engine or the coefficient of performance (COP) of a refrigerator, provide convenient measures for the performance of thermal machines. 
These figures are subject to universal bounds, which follow directly from the first and the second law of thermodynamics and are known as Carnot bounds \cite{Callen1985}. 
To attain its Carnot bound, a thermal machine has to work without producing any net entropy. 
This condition is generally assumed to be met only if the machine does not exchange any heat with its environment or if it operates infinitely slow.  
In both cases the generated output per time is zero. 
Hence, the Carnot limit can be reached only at the price of vanishing power.

The question how this trade-off can be formulated quantitatively for meso- and micro-scale thermal machines, and whether it can be overcome in special situations, has attracted significant interest over the last decade \cite{Benenti2011,Brandner2013,Allahverdyan2013,Holubec2016,Campisi2016,Polettini2017,Lee2017,Holubec2017,Holubec2018,Koyuk2019b,Miura2021}. 
As a result, a variety of trade-off relations that bound the power of different types of machines in terms of a dimensionless figure of merit were discovered, first in linear-response \cite{Brandner2015,Brandner2015b,Proesmans2015,Proesmans2016,Brandner2016} and then far from equilibrium \cite{Shiraishi2016,Pietzonka2018,Koyuk2019,Kamijima2021,Shiraishi2019,Menczel2020,Whitney2014,Whitney2015,Potanina2021,Miura2021b}. 
Since such bounds must go beyond the first and the second law, which due to the lack of a fundamental time scale do not provide any constraint on power, they have to be derived from the underlying dynamics of the system. 
As a result, different bounds hold for Markov jump processes \cite{Shiraishi2016,Pietzonka2018,Koyuk2019,Kamijima2021}, underdamped Fokker-Planck dynamics \cite{Shiraishi2016}, Lindblad dynamics \cite{Shiraishi2019,Menczel2020} or coherent transport \cite{Whitney2014,Whitney2015,Potanina2021}.  

For adiabatic thermal machines, which use a working system that is driven by slow periodic variations of external control parameters, thermodynamic geometry provides a promising avenue towards a unified picture. 
This framework was originally developed for macroscopic systems \cite{Weinhold1975,Gilmore1984,Andresen1988,Ruppeiner1995} and later extended to classical meso- and micro-scale systems \cite{Crooks2007,Sivak2012,Zulkowski2012,Matcha2015} as well as the quantum regime \cite{Brower1998,Scandi2019,Miller2019}. 
The key idea is that the time dependent state variables of the working system, e.g. the entries of a density matrix, which in general have to be found by solving a non-autonomous set of differential equations, become functions of the control variables and their time derivatives if the driving is slow on some characteristic time scale of the system. 
Quantities like work or entropy production, which depend on these state variables, can thus be related to geometric objects such as vector fields or metrics in the space of control parameters. 
Once the dynamics of the system has been specified, the coefficients defining these objects can be calculated by means of adiabatic perturbation theory \cite{Thomas2012,Ludovico2016,Cavina2017}.
Any relation between the quantities of interest, however, that follows from general symmetries or purely geometric arguments, holds universally for any kind of thermodynamically consistent dynamics. 

The geometric approach has lead to notable insights on the principles that govern the performance of adiabatic thermal machines \cite{Bustos2013,Izumida2021,Hino2021,Brandner2020,Miller2020,Miller2021,Frim2021,Hayakawa2021,Lu2022}.
Recent results include explicit optimization schemes for different types of devices \cite{Bhandari2020,Pancotti2020,Abiuso2020,Xu2021,Alonso2021,Watanabe2022} as well as geometric trade-off relations between the efficiency, power yield \cite{Brandner2020} and power fluctuations \cite{Miller2021} of cyclic heat engines that are driven by continuous temperature variations. 
These trade-off relations show that, close to the Carnot limit, the power of such devices is bounded by a linear function of their efficiency, which goes to zero at the Carnot value. 

In this article, we consider a complementary setting, where an adiabatic machine works between two thermal baths with fixed temperatures.
The thermodynamic geometry of this setup, which covers both heat engines and refrigerators, is usually developed by treating the temperature difference between the reservoir and the environment as a first-order perturbation along with the driving rates \cite{Bhandari2020}. 
Here, we argue that this approach is no longer sufficient if the machine operates close to the Carnot limit. 
Specifically, we show that, in this regime, the performance of a generic machine is governed by second-order corrections in the temperature difference between the baths. 
This effect leads to a new family of geometric trade-off relations implying a quadratic rather than a linear decay of power at the Carnot bound. 

\newcommand{\bl}{\boldsymbol{\lambda}}
\renewcommand{\l}{\lambda}
\renewcommand{\tint}{\int_0^\tau\!\! dt \;}
\renewcommand{\b}{\beta}
\newcommand{\be}{\beta_\text{e}}
\newcommand{\br}{\beta_\text{r}}

This behavior can be derived from a general scaling argument. 
To this end, it is convenient to introduce generalized fluxes $J_x$ and affinities $A_x$ such that the average rate of entropy production can be expressed in the standard form $\sigma = J_x A_x$, where summation over identical indices is understood throughout \cite{Seifert2012}. 
The fluxes $J_x$ correspond to output and input of the machine and the affinities $A_x$ represent the thermodynamic forces that drive the system away from equilibrium. 
For an adiabatic-response theory, natural choices of these variables are
\begin{equation}
J_w=W, \quad J_q=Q/\tau, \;\;\; A_w=\be/\tau\;\;\; \text{and}\;\;\; A_q=\be-\br. 
\end{equation}
Here, $\tau$ denotes the cycle time, $\be$ and $\br$ are the inverse temperatures of the two baths, to which we refer as environment and reservoir; $W$ and $Q$ are the applied work and the heat uptake from the reservoir per operation cycle. 
Boltzmann's constant is set to 1 throughout. 

\newcommand{\ve}{\varepsilon}
\renewcommand{\vec}{\varepsilon_\text{C}}

We now focus on refrigerators. 
That is, we assume that $\br\geq \be$ and $W, Q\geq 0$ so that the machine absorbs work from the external driving and extracts heat from the cold reservoir.
The performance of such a device is described by the COP $\ve\equiv Q/W$, which is bounded by the Carnot value $\vec\equiv \be/(\br-\be)$. 
To determine under what conditions $\ve$ approaches $\vec$, we divide the work input into an isothermal part and a correction stemming from the temperature difference between the reservoirs, 
\begin{equation}\label{Eq:FNLw}
J^\text{iso}_w \equiv J_w|_{A_q=0}\equiv K_{ww}A_w
    \quad\text{and}\quad
    J_w -J_w^\text{iso}\equiv K_{wq} A_q. 
\end{equation}
Analogously, the heat flux can be divided into a quasi-static contribution and a finite-rate correction,
\begin{equation}\label{Eq:FNLq}
    J_q^\text{qs}\equiv J_q|_{A_w=0}\equiv K_{qq} A_q
    \quad\text{and}\quad
    J_q -J_q^\text{qs} \equiv K_{qw}A_w. 
\end{equation} 
The coefficients $K_{xy}$ are functions of the affinities, which in general assume finite values in the limit $A_x\rightarrow 0$.
Furthermore, the second law requires that $K_{ww}, K_{qq} \geq 0$ and time-reversal symmetry implies that the cross-coefficients obey the Onsager symmetry 
\begin{equation}\label{Eq:OS}
    K_{qw}|_{A_x=0} = - K_{wq}|_{A_x=0}
\end{equation}
in zeroth order with respect to the affinities. 
This symmetry holds for arbitrary driving protocols as long as the system is not subject to external magnetic fields breaking time reversal symmetry, which we assume here. 

The normalized COP can now be written in the form
\begin{equation}\label{Eq:NCOP}
    \frac{\ve}{\vec} = -\frac{J_q A_q}{J_w A_w} = -\frac{K_{qw} + K_{qq}(A_q/A_w)}{K_{wq} + K_{ww}(A_w/A_q)}. 
\end{equation}
Since the isothermal work will in general not vanish, it is natural to assume that $K_{ww}>0$. 
Owing to the Onsager symmetry \eqref{Eq:OS}, the expression \eqref{Eq:NCOP} then converges to $1$ in the quasi-static limit $A_w\rightarrow 0$ if $K_{qq}=0$ and $A_q\propto A_w^{\alpha}$ with $0<\alpha<1$. 
That is, provided that the quasi-static heat flux vanishes, the Carnot bound is attained asymptotically as both affinities go to zero with $A_w$ vanishing faster than $A_q$, whereby both $\ve$ and $\vec$ diverge. 

To determine how the cooling power $J_q$ decays in this limit, we expand the coefficients $K_{qw}$ and $K_{wq}$ in the affinities keeping leading and first sub-leading terms,
\begin{equation}\label{Eq:K1Order}
    K_{qw} = L_{qw}+ L^q_{qw}A_q, \quad
    K_{wq} = -L_{qw} + L^q_{wq}A_q. 
\end{equation}
Inserting these expansions into Eq.~\eqref{Eq:NCOP} and again keeping only leading and first subleading terms leaves us with
\begin{equation}\label{Eq:COPExp}
\frac{\ve}{\vec}= 1 +\frac{L^q_{qw}+L^q_{wq}}{L_{qw}}A_q + \frac{L_{ww}}{L_{qw}}\frac{A_w}{A_q},
\end{equation}
where $L_{ww}\equiv K_{ww}|_{A_x=0}$. 
Upon maximizing the right-hand side of this equation with respect to $A_q$, we obtain an upper bound on $\ve/\vec$ and an optimum for the thermal gradient, which are given by 
\begin{equation}\label{Eq:RCOPBnd}
    \ve/\vec\leq 1- \sqrt{L_{qw} A_w/Z} \quad\text{and}\quad 
    A_q^\ast = -\sqrt{z A_w}
\end{equation}
with $Z\equiv L_{qw}^3/4(L^q_{wq}+L^q_{qw})L_{ww}$ and $z\equiv L_{ww}/(L^q_{wq}+L^q_{qw})$ being non-negative quantities \footnote{To see that $Z,z\geq 0$, first notice that the refrigerator condition $J_q\geq 0$ requires that $L_{qw}\geq 0$, since $J_q = L_{qw}A_w$ in leading order in the affinities and $A_w\geq \mathcal{0}$
Second, observe that the rate of entropy production is given by $\sigma = L_{ww}A_w^2+(L^q_{qw}+L^q_{wq})A_wA_q^2$ at leading order. 
The second law $\sigma\geq 0$ thus requires $L^q_{qw}+L^q_{wq}\geq 0$. 
}. 
Since $J_q= L_{qw}A_w$ in leading order, we can now replace $A_w$ with $J_q/L_{qw}$ in Eq.~\eqref{Eq:RCOPBnd}, which yields the 
power-COP trade-off relation
\begin{equation}\label{Eq:RTOR}
    J_q \leq Z (\vec-\ve)^2/\vec^2. 
\end{equation}
This relation, which is our first main result, shows that the cooling power of a generic adiabatic refrigerator decays at least quadratically at the Carnot bound.

\newcommand{\etac}{\eta_\text{C}}
A similar picture emerges for adiabatic heat engines, which are realized for $\br\leq \be$, $Q\geq 0$ and $W\leq 0$. 
Hence, the machine picks up heat from the hot reservoir and generates work output. 
Its efficiency is then defined as $\eta\equiv - W/Q$ and the corresponding Carnot bound reads $\etac\equiv (\be-\br)/\be$. 
Upon introducing the normalized efficiency $\eta/\etac= - J_w A_w/J_q A_q$, the steps that lead to Eq.~\eqref{Eq:RTOR} can be repeated one by one \cite{SM}. 
We thus find that $\eta/\etac$ generically converges to $1$ only if the quasi-static heat flux vanishes and both affinities go to zero with $A_w$ vanishing faster than $A_q$, whereby $\eta$ and $\etac$ both approach zero. 
Close to this limit, the engine is subject to the power-efficiency trade-off relation
\begin{equation}\label{Eq:HETOR}
    P \leq Z(\etac-\eta)^2/\etac
\end{equation}
and the optimal thermal gradient, for which it is saturated asymptotically, is given by $A_q^\ast=\sqrt{z A_w}$. 

The bounds \eqref{Eq:RTOR} and \eqref{Eq:HETOR} ultimately arise from the fact that the Onsager symmetry \eqref{Eq:OS} does not extend to the second-order coefficients $L^q_{wq}$ and $L^q_{qw}$. 
Still, there are special situations, where $L^q_{wq}\simeq - L^q_{qw}$ \cite{SM}. 
Under this condition, the second term in the expansion \eqref{Eq:COPExp} can be neglected and we are left with the trivial relation $\ve = (1+ L_{ww} A_w/L_{qw}A_q)\vec$ \footnote{
For heat engines, the symmetry $L_{wq}\simeq -L_{qw}$ leads to the analogous relation $\eta = (1-L_{ww} A_w/L_{qw}A_q)\etac$.}. 
The Carnot bound is then attained for any $A_q$ in the limit $A_w\rightarrow 0$ with the power of the machine vanishing linearly.
However, this behavior will typically occur only in fine-tuned systems. 
We stress that this restriction appears only when sub-leading terms in the expansions of the generalized fluxes are taken into account, cf. Eq.~\eqref{Eq:ARExpFl}. 
It is therefore not captured by the established adiabatic-response theory, where both fluxes are assumed to be linear in affinities \cite{Bhandari2020}. 

To unveil the geometric character of the bounds \eqref{Eq:RTOR} and \eqref{Eq:HETOR}, we have to analyze the structure of $Z$. 
We assume that the machine is driven by periodic changes of the parameters $\bl=\{\l^\mu\}$, which control the energy of the working system and its coupling to the baths. 
Once the system has settled to a periodic state, the work input and heat uptake from the reservoir per cycle are given by 
\begin{equation}\label{Eq:WQ}
    W= -\tint f^\mu_t \dot{\l}^\mu_t
    \quad\text{and}\quad
    Q= \tint j_t,
\end{equation}
where $f^\mu_t$ is the generalized force conjugate to the parameter $\l^\mu$ and $j_t$ denotes the heat current from the reservoir into the system. 
If the driving is slow on the internal time-scale of the system and the difference between the inverse temperatures of reservoir and environment is small on its typical energy scale, these quantities can be expanded in the driving rates and the thermal gradient, 
\begin{subequations}\label{Eq:ARExp}
\begin{align}
\label{Eq:ARExpf}
f^\mu_t &=-\partial_\mu \mathcal{F}_{\bl_t} - \mathcal{R}^{\mu\nu}_{\bl_t}\be\dot{\l}^\nu_t 
        - \mathcal{R}^{\mu q}_{\bl_t}A_q - \mathcal{R}^{\mu qq}_{\bl_t} A_q^2,\\
j_t &= \mathcal{R}^{q\mu}_{\bl_t}\be\dot{\l}^\mu_t  + \mathcal{R}^{qq\mu}_{\bl_t}\be\dot{\l}_t^\mu  A_q. 
\end{align}
\end{subequations}
The free energy of the system $\mathcal{F}_{\bl}$ and the adiabatic-response coefficients $\mathcal{R}_{\bl}$ depend parametrically on the control vector $\bl$ and on $\be$. 
Note that we include only the relevant second-order terms and assume that there are no heat-leaks, i.e. $j_t|_{\dot{\bl}_t=0}=0$. 

Upon inserting the Eqs.~\eqref{Eq:ARExp} into Eq.~\eqref{Eq:WQ} and comparing the result with the expansions of the fluxes, 
\begin{equation}\label{Eq:ARExpFl}
    J_w = L_{wx}A_x + L^q_{wq}A_q^2, \quad
    J_q = L_{qw}A_w + L^q_{qw}A_w A_q,
\end{equation}
the off-diagonal coefficients can be expressed as line integrals in the space of control parameters, 
\begin{equation}\label{Eq:KCARC}
    \left[\begin{array}{ll}
    L_{wq}     &  L_{qw}\\
    L^q_{wq}    & L^q_{qw}
    \end{array}\right]
    = \oint_\gamma \;
    \left[\begin{array}{ll}
    \mathcal{A}^{\mu q}_{\bl}     &  \mathcal{A}^{q\mu}_{\bl} \\[3pt]
    \mathcal{A}^{\mu qq}_{\bl}     &  \mathcal{A}^{qq\mu}_{\bl} 
    \end{array}\right] \; d\l^\mu. 
\end{equation}
Here, $\gamma$ denotes the closed path that is mapped out by the driving protocols $\bl_t$ and the thermodynamic vector potentials are defined as 
\begin{equation}
    \left[\begin{array}{ll}
    \mathcal{A}^{\mu q}_{\bl}     &  \mathcal{A}^{q\mu}_{\bl} \\[3pt]
    \mathcal{A}^{\mu qq}_{\bl}     &  \mathcal{A}^{qq\mu}_{\bl} 
    \end{array}\right]
    \equiv
   -\l^\nu\partial_\mu \left[\begin{array}{ll}
    \mathcal{R}^{\nu q}_{\bl}     &  \mathcal{R}^{q\nu}_{\bl} \\[3pt]
    \mathcal{R}^{\nu qq}_{\bl}     &  \mathcal{R}^{qq\nu}_{\bl} 
    \end{array}\right]
\end{equation}
The coefficient $L_{ww}$ does not admit a geometric representation. 
It is, however, subject to the geometric bound
\begin{equation}\label{Eq:TL}
    L_{ww} = \tau \tint \mathcal{G}^{\mu\nu}_{\bl_t} \dot{\l}^\mu_t\dot{\l}^\nu_t\geq \mathcal{L}^2 \;\;\text{with}\;\;
    \mathcal{L}\equiv\oint_\gamma \sqrt{\mathcal{G}^{\mu\nu}_{\bl}d\l^\mu d\l^\nu}
\end{equation}
being the thermodynamic length of the path $\gamma$. 
This notion is motivated by the fact that, owing to the second law, the coefficients $\mathcal{G}^{\mu\nu}_{\bl}\equiv (\mathcal{R}^{\mu\nu}_{\bl}+\mathcal{R}^{\nu\mu}_{\bl})/2$ form a positive semi-definite matrix and can therefore be interpreted as a pseudo-Riemannian metric in the space of control parameters. 
The bound \eqref{Eq:TL} can be derived by minimizing $L_{ww}$ with respect to the parameterization of the path $\gamma$. 
The optimal parameterization $\phi_t$, for which Eq.~\eqref{Eq:TL} becomes an equality, is implicitly determined by the condition 
\begin{equation}\label{Eq:SF}
    t = \frac{\tau}{\mathcal{L}} \int_0^{\phi_t} \! ds \; \sqrt{\mathcal{G}^{\mu\nu}_{\bl_s}\dot{\l}^\mu_s \dot{\l}^\nu_s}. 
\end{equation}

The Eqs.~\eqref{Eq:KCARC} and \eqref{Eq:TL} show that the figure of merit $Z$ is subject to the bound $Z\leq\mathcal{Z}\equiv L_{qw}^3/4(L^q_{wq}+L^q_{qw})\mathcal{L}^2$, where $\mathcal{Z}$ depends only on geometric quantities. 
Thus, Eqs.~\eqref{Eq:RTOR} and \eqref{Eq:HETOR} imply the geometric trade-off relations 
\begin{equation}\label{Eq:GTOR}
    J_q \leq \mathcal{Z}(\vec-\ve)^2/\vec^2 \quad\text{and}\quad P\leq \mathcal{Z}(\etac-\eta)^2/\etac
\end{equation}
for adiabatic refrigerators and heat engines. 
These bounds, which are our second main result, hold for any thermodynamically consistent dynamics that admits a well-defined adiabatic-response regime. 
Moreover, they are asymptotically saturated in the limit $A_w\rightarrow 0$ if the optimal parameterization $\phi_t$ is chosen for the control path $\gamma$ and the thermal gradient is scales with the driving frequency as $A_q=\mp\sqrt{\hat{z} A_w}$ with $\hat{z}\equiv \mathcal{L}^2/(L^q_{wq}+L^q_{qw})$. 

Two-stroke cycles provide a general mechanism to fully suppress heat-leaks. 
Under this protocol, the working system is decoupled from the environment for the first part $\tau_1<\tau$ of the cycle and decoupled from the reservoir during the second part $\tau-\tau_1$. 
As a result, no persistent heat current between reservoir and environment emerges and $j_t$ and $J_q$ vanish for $A_w\rightarrow 0$. 
The coefficients \eqref{Eq:KCARC} then depend solely on equilibrium properties of the working system and the 
geometric figure of merit becomes 
\begin{equation}\label{Eq:ZTS}
    \mathcal{Z}= \frac{(\mathcal{S}_{\bl_1}-\mathcal{S}_{\bl_0})^3}{2\be^3(\mathcal{C}_{\bl_1}+\mathcal{C}_{\bl_0})\mathcal{L}^2}, 
\end{equation}
where $\bl_1\equiv\bl_{\tau_1}$ and $\mathcal{S}_{\bl}$ and $\mathcal{C}_{\bl}$ denote the equilibrium entropy and heat capacity of the working system at fixed control parameters and inverse temperature $\be$ \cite{SM}. 
Hence, the only quantity that still depends on the dynamics of the system is the thermodynamic length $\mathcal{L}$.

\begin{figure}
\includegraphics[scale=0.89]{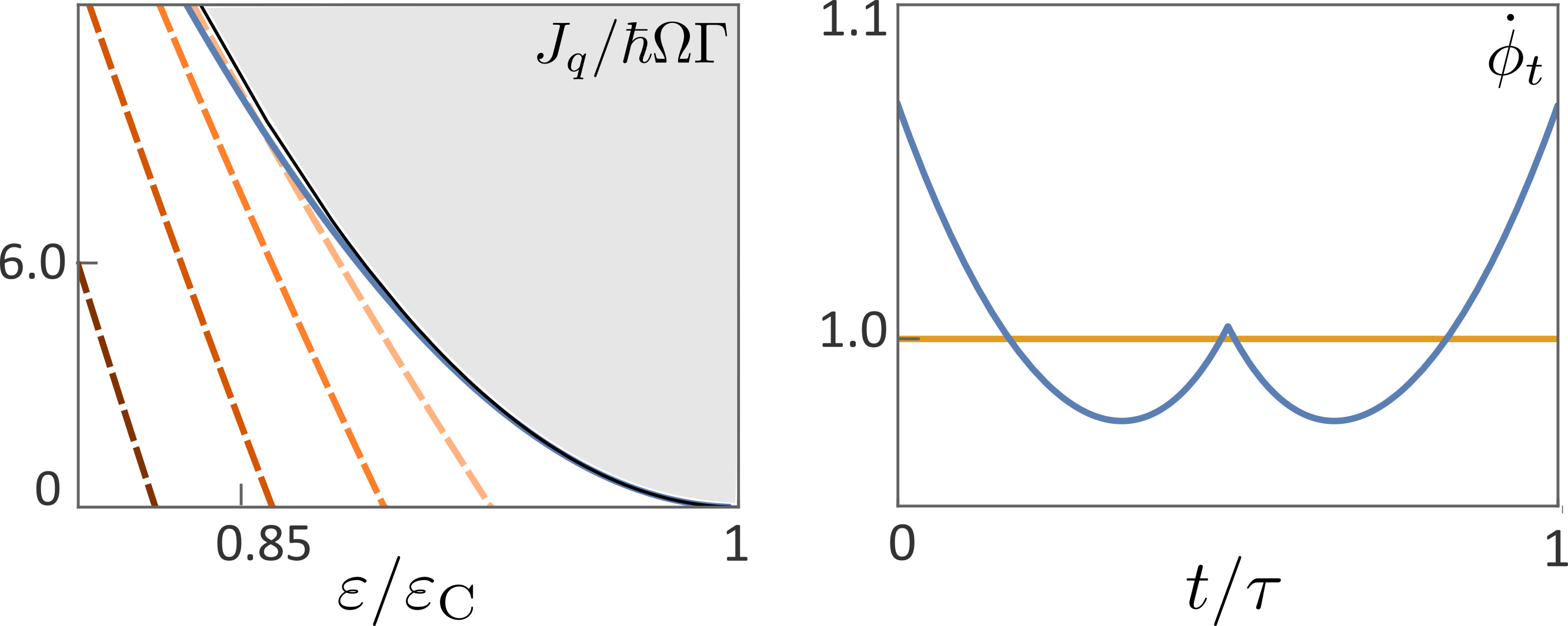}
\caption{Qubit refrigerator. 
The first plot shows the cooling power in units of $10^{-4}\hbar\Omega\Gamma$ as a function of the normalized COP, where $\tau$ is varied from $10^{-6}/\Gamma$ to $1/\Gamma$. 
Along the dashed lines, for which $1/\br=0.89,...,0.95\hbar\Omega$ from left to right, the cooling power goes to zero at $\ve/\vec<1$ in the limit $A_w\rightarrow 0$. 
The solid line, for which we set $A_q= -\sqrt{\hat{z} A_w}$, almost saturates the first bound \eqref{Eq:GTOR}. 
For all plots, we have set $1/\be=\hbar\Omega$ and chosen the optimal parameterization of the control path $\phi_t$, whose derivative is shown in the second plot. 
\label{Fig:TLR}}
\end{figure}

To illustrate the two-stroke mechanism, we consider a quantum refrigerator that consists of a qubit with Hamiltonian $H_{\l}=\hbar\Omega\l\sigma_z/2$, where $\Omega$ sets the energy scale \cite{Juka2007,Juka2016}. 
The state $\rho_t$ of the system evolves according to the adiabatic Lindblad equation \cite{Albash2012}
\begin{equation}\label{Eq:ME}
\partial_t \rho_t = -\frac{i}{\hbar}[H_{\l_t},\rho_t]
         +\mathsf{D}_{\bl_t}^\text{r}\rho_t + \mathsf{D}_{\bl_t}^\text{e}\rho_t
\end{equation}
with
$
\mathsf{D}_{\bl}^\text{x}\cdots \equiv 
    \Gamma \kappa^\text{x}_t \sum\nolimits_{\alpha=\pm} n^\alpha_{\l,\b_\text{x}}\bigl([\sigma_\alpha \cdots,\sigma_\alpha^\dagger] + [\sigma_\alpha,\cdots \sigma_\alpha^\dagger]\bigr)
$ 
being dissipation super-operators that describe the influence of the reservoir and the environment ($\text{x}=\text{r},\text{e}$). 
Here, $\sigma_z$ and $\sigma_\pm$ are the usual Pauli matrices, the rate $\Gamma>0$ sets the relaxation time scale of the system and $n^+_{\l,\b_\text{x}}\equiv 1/(e^{\b_\text{x}\hbar\Omega\l}-1)$ and $n^-_{\l,\b_\text{x}}\equiv n^+_{\l,\b_\text{x}}+1$ are thermal factors. 
For the control parameters $\l^1\equiv \l$ and $\l^2\equiv \kappa^\text{r}\equiv 1- \kappa^\text{e}$, we choose the following protocols. 
During the first stroke, the system couples to the reservoir, i.e. $\kappa^\text{r}_t=1$, and the level spacing $\l_t$ decreases linearly from $2$ to $1$; in the second stroke, the system couples to the environment, i.e. $\kappa^\text{r}_t=0$, and $\l_t$ increases linearly from $1$ to $2$. 
\newcommand{\tr}[1]{\text{tr}[#1]}
For general adiabatic Lindblad dynamics, the generalized forces and the heat current are given by
\begin{equation}
    f_t^\mu = -\tr{\rho^\tau_t \partial_\mu H_{\bl_t}} \quad\text{and}\quad
    j_t = \tr{(\mathsf{D}^\text{r}_{\bl_t}\rho^\tau_t) H_{\bl_t}},
    \label{Eq: Lin f j}
\end{equation}
where $\rho^\tau_t$ is the periodic state of the system. 
These quantities can now be calculated perturbatively in the driving rates and the thermal gradient, which yields the thermodynamic length \eqref{Eq:TL} and the figure of merit \eqref{Eq:ZTS} for the qubit refrigerator \cite{SM}.
To compare the performance of this device with the first trade-off relation \eqref{Eq:GTOR}, we calculate its COP and cooling power by solving the master equation \eqref{Eq:ME} numerically.
Fig.~\ref{Fig:TLR} shows that, in the quasi-static limit, $\ve$ remains indeed strictly smaller than $\vec$ for any fixed $A_q<0$, while it approaches  approaches $\vec$ if $A_q$ is optimized with respect to the cycle time;
the bound \eqref{Eq:GTOR} is asymptotically saturated if the optimal parameterization \eqref{Eq:SF} is chosen for the control path. 

\begin{figure}
\includegraphics[scale=0.85]{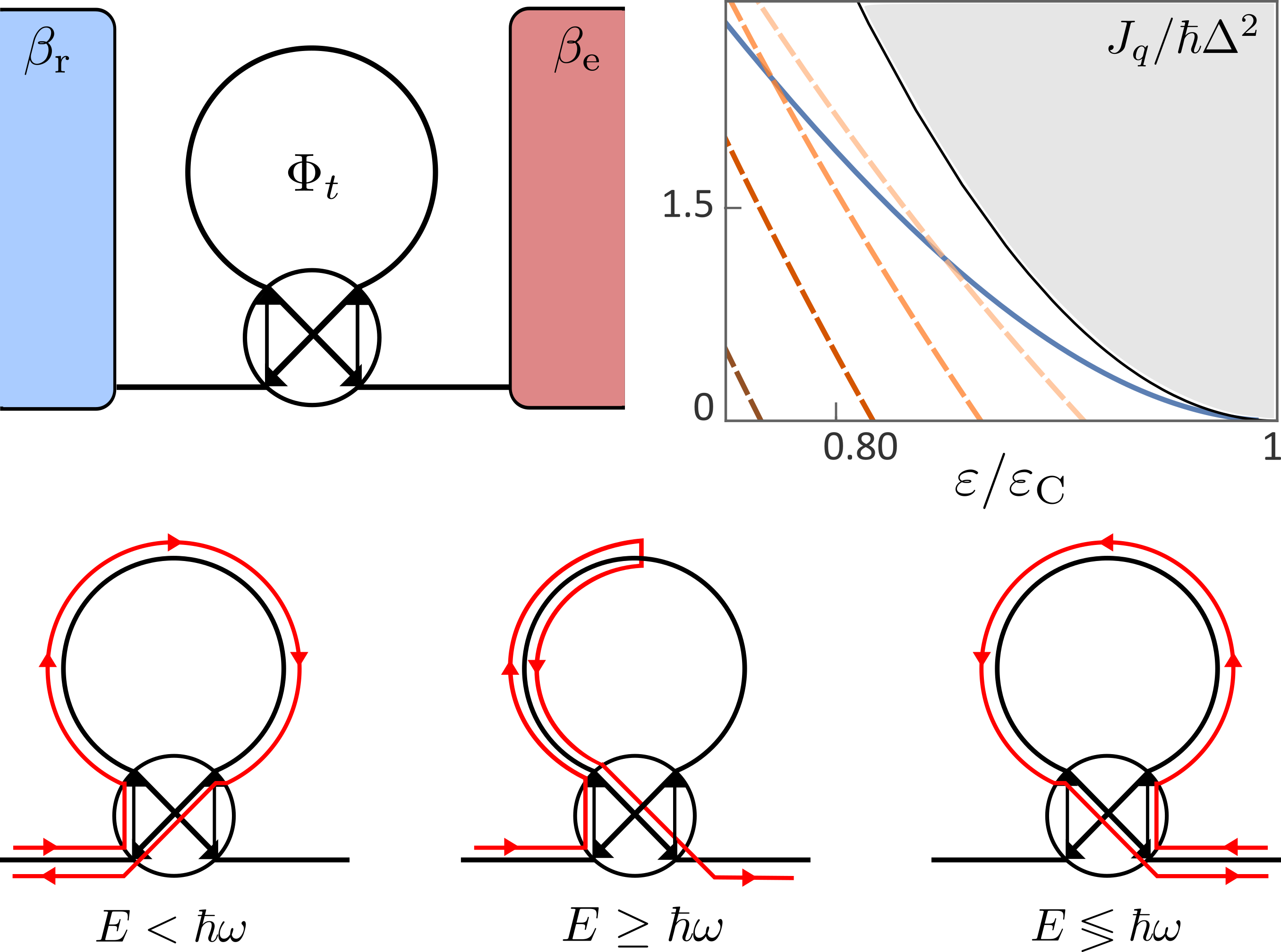}
\caption{Aharonov-Bohm refrigerator. 
A four-way beam splitter connects a mesoscopic ring and two ideal leads supporting a single transport channel. 
The left and the right lead are coupled to the cold reservoir and the environment, respectively. 
The linearly increasing magnetic flux $\Phi_t$ induces a constant electromagnetic force around the ring, which decelerates incoming carriers from the reservoir; 
carriers with energy $E\geq \hbar\omega$ pass through the ring and return to the reservoir; carriers with $E<\hbar\omega$ are reflected on the ring and transmitted to the environment. 
Incoming carriers from the environment are accelerated and return to the environment regardless of their energy. 
For $\omega\rightarrow 0$, no carriers are transmitted; hence, the quasi-static heat flux vanishes.
The plot shows the cooling power of the device in units of $10^{-4}\hbar \Delta^2$ as a function of its normalized COP, where $\omega$ is varied from $10^{-7}\Delta$ to $10^{-2}\Delta$. 
Here, $2\pi/\Delta =4\pi m l^2/\hbar$ is the typical dwell time, $m$ denotes the carrier mass and $l$ the circumference of the ring. 
The dashed lines correspond to $1/\br =0.90,...,0.96\hbar\Delta$ from left to right. 
The blue line is obtained for $A_q=-\sqrt{\hat{z}A_w}$ and the gray area indicates the bound \eqref{Eq:GTOR}. 
For all plots, we have set $1/\be = \hbar\Delta$ and $\mu=1.05\hbar\Delta$ and used the optimal parameterization $\phi_t =t$. 
\label{Fig:AB}
}
\end{figure}

This result shows that the sudden changes of the coupling parameters $\kappa^\text{r}$ and $\kappa^\text{e}$ are consistent with the adiabatic-response assumption. 
This conclusion holds in general for the weak-coupling regime, where the internal energy and the equilibrium state of the system do not depend on its interaction with the baths. 
Under this condition, the coupling parameters do not give rise to generalized forces and their time derivatives do not appear in the expansion \eqref{Eq:ARExpf}, see \cite{SM} for details. 
As a result, the coupling parameters do enter the diagonal kinetic coefficient \eqref{Eq:TL} or the thermodynamic length. 
They rather affect only the off-diagonal coefficients \eqref{Eq:KCARC}, which, being geometric quantities, do not depend on the driving rates. 
The expansions \eqref{Eq:ARExpFl} of the generalized fluxes are thus well-defined for the two-stroke protocol. 

\newcommand{\ket}[1]{|#1\rangle}
\newcommand{\bra}[1]{\langle #1|}
To show that the trade-off relations \eqref{Eq:GTOR} are applicable also outside the two-stroke scheme, we now consider a mesoscopic refrigerator based on coherent transport. 
The system consists of a four-way beam-splitter and a mesoscopic ring subject to the time-dependent magnetic flux $\Phi_t$, see Fig.~\ref{Fig:AB} \cite{Avron2001}.  
Its two control parameters can be identified with the real and the imaginary part of the Aharonov-Bohm phase that is picked by a particle when passing through the ring, i.e. $e^{iq \Phi/\hbar c}\equiv\lambda^1 + i\lambda^2$, where $c$ is the speed of light and $q$ the carrier charge. 
A linearly increasing flux, $\Phi_t \equiv \hbar c \omega t/ q$, thus leads to the driving protocols $\l^1_t = \cos[\omega t]$ and $\l^2_t = \sin[\omega t]$, where $\omega = 2\pi/\tau$. 

For coherent transport and effectively non-interacting carriers, the generalized forces and the heat current admit the general expressions \cite{Bode2012,Thomas2012,Brandner2020b}
\begin{subequations}
\begin{align}
    f^\mu_t &= - \int_0^\infty \!\!\! dE \sum_{\text{x}=\text{r},\text{e}}\bra{\psi_{E,t}^{\text{x}}}\partial_\mu H_{\bl_t}\ket{\psi_{E,t}^{\text{x}}} g^\text{x}_E\quad\text{and}\\
    j_t &= \int_0^\infty \!\!\! dE\sum_{\text{x}=\text{r},\text{e}} \bra{\psi_{E,t}^{\text{x}}}J_{\bl_t}\ket{\psi_{E,t}^{\text{x}}} g^\text{x}_E. 
\end{align}
\end{subequations}
Here, $H_{\bl}$ and $J_{\bl}$ are the single-carrier Hamiltonian and heat current operator and $\ket{\psi^\text{x}_{E,t}}$ is the 
Floquet scattering state that describes a carrier with energy $E$, which enters the system either from the reservoir or the environment \cite{Moskalets2002,Brandner2020b};  
$g^\text{x}_E\equiv 1/[1+e^{\beta_\text{x}(E-\mu)}]$ is the Fermi function with chemical potential $\mu$. 

If the typical dwell time of carriers inside the system is short compared to $\tau$, the Floquet-scattering states can be calculated perturbatively \cite{Thomas2012}, which yields the figure of merit $\mathcal{Z}$ for the Aharonov-Bohm refrigerator \cite{SM}. 
Since no carriers are transmitted for $\omega =0$, the quasi-static heat flux vanishes and the first trade-off relation \eqref{Eq:GTOR} applies. 
Fig.~\ref{Fig:AB} shows how the cooling power and the COP of the Aharonov-Bohm refrigerator, which can be calculated exactly \cite{SM}, compare to this bound in the slow-driving regime. 
As for the qubit-refrigerator, we find that $\ve$ does not reach $\vec$ for any fixed $A_q<0$, while the trade-off relation \eqref{Eq:GTOR} is asymptotically saturated in the quasi-static limit if $A_q$ is optimized with respect $A_w$.

This outcome further underlines the universality of our main insights. 
First, generic adiabatic thermal machines cannot approach their Carnot limit when working between two baths with finite temperature difference.
Second, close to this limit, the performance of such devices is not captured by standard adiabatic-response theory, which treats both temperature gradient and driving rates as first-order perturbations. 
Instead, second-order terms describing corrections to the finite-rate heat current and the non-isothermal work play an essential role. 
Taking these corrections into account leads to the geometric trade-off relations \eqref{Eq:GTOR}, which imply that power decays quadratically rather than linear at the Carnot bound. 
These results follow only from system-independent arguments and the Onsager symmetry \eqref{Eq:OS}. 
It now remains to future research to explore how breaking this symmetry alters the performance of adiabatic thermal machines. 

\begin{acknowledgments}
\emph{Acknowledgments:}
The authors thank Thomas Veness for a careful proofreading of this manuscript. 
K.B. acknowledges support from the University of Nottingham through a Nottingham Research Fellowship and from UK Research and Innovation through a Future Leaders Fellowship (Grant Reference: MR/S034714/1).
\end{acknowledgments}

\end{document}



\title{Geometric Bounds on the Power of Adiabatic Thermal Machines\\ - Supplemental Material -}
\author{Joshua Eglinton}
\author{Kay Brandner}
\affiliation{
School of Physics and Astronomy,
University of Nottingham,
Nottingham NG7 2RD,
United Kingdom\\
Centre for the Mathematics and Theoretical Physics of Quantum Non-equilibrium Systems,
University of Nottingham,
Nottingham NG7 2RD, 
United Kingdom
}
\date{\today}
\maketitle
\renewcommand{\theequation}{S\arabic{equation}}
\makeatletter
\newcommand{\manuallabel}[2]{\def\@currentlabel{#2}\label{#1}}
\makeatother

\manuallabel{MT:FluxAff}{1}
\manuallabel{MT:OS}{4}
\manuallabel{MT:K1Order}{6} 
\manuallabel{MT:COPExp}{7}
\manuallabel{MT:RCOPBnd}{8} 
\manuallabel{MT:RTOR}{9}
\manuallabel{MT:HETOR}{10} 
\manuallabel{MT:WQ}{11}
\manuallabel{MT:ARExpf}{12}
\manuallabel{MT:FluxExp}{13}
\manuallabel{MT:TL}{16}
\manuallabel{MT:ZTS}{19}
\manuallabel{MT:Lin f j}{21}
\manuallabel{MT:SF}{17}

\newcommand{\etac}{\eta_\text{C}}
\newcommand{\be}{\beta_{\text{e}}}
\newcommand{\br}{\beta_{\text{r}}}
\newcommand{\at}{\alpha_{t}}

\newcommand{\bl}{\boldsymbol{\lambda}}
\renewcommand{\l}{\lambda}
\renewcommand{\tint}{\int_0^\tau\!\! dt \;}
\renewcommand{\b}{\beta}
\newcommand{\tr}[1]{\text{tr}[#1]}

\section{Preamble}
This supplemental material consists of three parts: 
In Sec.~II, we derive the power-efficiency trade-off relation for adiabatic heat engines that was discussed in the main text. 
In Sec.~III, we provide general expressions for the kinetic coefficients of two-stroke cycles along with a detailed discussion of the qubit refrigerator. 
In Sec.~VI, we sketch the derivation of the kinetic coefficients for periodically driven coherent conductors and further discuss the Aharonov-Bohm refrigerator. 
All symbols are used with same definitions as in the main text. 
Equation numbers without additional indicator S refer to the main text. 

\section{Power-Efficiency Trade-off Relation for Adiabatic Heat Engines}\label{sec: 1}
The power-efficiency trade-off relation \eqref{MT:HETOR} can be derived by following the same steps that lead to the power-COP trade-off relation \eqref{MT:RTOR} in the main text. 
We first express the normalized efficiency as
\begin{equation}\label{Eq:X}
    \frac{\eta}{\etac} = -\frac{J_{w}A_{w}}{J_{q}A_{q}}=-\frac{K_{ww}(A_{w}/A_{q})+K_{wq}}{K_{qq}(A_{q}/A_{w})+K_{qw}}. 
\end{equation}
Assuming that the isothermal work does not vanish, i.e. $K_{ww}>0$, this expression converges to $1$ in the quasi-static limit $A_{w}\rightarrow 0$ if $K_{qq}=0$; 
this observation follows directly from the Onsager symmetry \eqref{MT:OS}. 
To determine, how the power of the engine decays in this limit, we insert the expansions \eqref{MT:K1Order} for the coefficients $K_{wq}$ and $K_{qw}$ into Eq.~\eqref{Eq:X}. 
Keeping only leading and first sub-leading contributions yields 
\begin{equation}\label{Eq:SMCOPEx}
    \frac{\eta}{\eta_{C}}=1-\frac{L_{qw}^{q}+L_{wq}^{w}}{L_{qw}}A_{q}-\frac{L_{ww}}{L_{qw}}\frac{A_{w}}{A_{q}}.
\end{equation}
Maximising this expression with respect to $A_{q}$ returns the results 
\begin{equation}
    \eta/\eta_{C} \leq 1-\sqrt{L_{qw} A_w/Z}\quad\text{and}\quad A_{q}^{\ast}=\sqrt{zA_{w}},
    \label{SM: HEeffBnd}
\end{equation}
for the upper bound on the normalized efficiency and the optimum of $A_{q}$.
Note that the condition $J_w<0$ requires $L_{wq}<0$ and thus $L_{qw}>0$. 
Finally, we observe that the power output of a heat engine, $P=-W/\tau=-A_wJ_w/\be$, reduces to $P=-L_{wq} A_w A_q/\be$ in leading order in the affinities. 
Thus, we can eliminate $A_w$ in favor of $P$ in the bound \eqref{Eq:X}, which leaves us with the result 
\begin{equation}
    P\leq Z(\etac-\eta)^2/\etac. 
\end{equation}
Here, we have used that $L_{wq}=-L_{qw}$ and $\etac=A_q/\be$. 

\section{Two-Stroke Cycles}
In the following, we derive microscopic expressions for the adiabatic-response coefficients describing general two-stroke cycles and calculate these coefficients explicitly for the qubit-refrigerator discussed in the main text. 

\subsection{Kinetic Coefficients}
We consider a thermal machine, whose working system is described by the Hamiltonian $H_{\bl}$. 
The system is driven into a periodic state $\rho_t^\tau$ through cyclic variations of the control parameters $\bl=\{\l^\mu\}$. 
During the first stroke of the cycle, where $0\leq t < \tau_1$, the system is coupled to a reservoir with inverse temperature $\br$; during the second stroke, where $\tau_1\leq t <\tau$, it couples to an environment with inverse temperature $\be$. 
The energy balance of the machine is described by the first law 
\begin{equation}\label{SM:FL}
\dot{E}_t = -f^\mu_t \dot{\lambda}_t^\mu + j_t
\end{equation}
where $E_t = \tr{\rho^\tau_t H_{\bl_t}}$ denotes the internal energy of the working system. 
The generalized forces and the rate of heat uptake are given by 
\begin{equation}\label{SM:Deffj}
    f^\mu_t = -\tr{\rho^\tau_t\partial_\mu H_{\bl_t}} \quad\text{and}\quad
    j_t = \tr{\dot{\rho}^\tau_t H_{\bl_t}}. 
\end{equation}
Hence, the applied work in the first and the second stroke can be expressed as 
\begin{equation}\label{SM:TSwork}
    W_1 = - \int_0^{\tau_1} \! dt \; f^\mu_t \dot{\l}^\mu_t 
    \quad\text{and}\quad
    W_2 = - \int_{\tau_1}^{\tau} \! dt \; f^\mu_t \dot{\l}^\mu_t.
\end{equation}

In the adiabatic-response regime, the generalized forces can be expanded as 
\begin{equation}\label{SM:fExp}
    f^\mu_t = -\partial_\mu\mathcal{F}^\text{x}_{\bl_t}-\be\mathcal{R}^{\mu\nu}_{\bl_{t}}\dot{\lambda}^{\nu}_t,
\end{equation}
where $\mathcal{F}^\text{x}_{\bl}$ denotes the equilibrium free energy of the system at the inverse temperature $\beta_\text{x}$ with $\text{x}=\text{r}$ during the first stroke and $\text{x}=\text{e}$ during the second stroke. 
The adiabatic-response coefficient $\mathcal{R}^{\mu\nu}_{\bl}$ depend only on $\be$, since we neglect cross terms between the driving rates and the thermal gradient $A_q= \be - \br$. 
Note that Eq.~\eqref{SM:fExp} follows directly from the definition of the generalized forces \eqref{SM:Deffj} and the assumption that the system follows its Gibbs state $\varrho^\text{x}_{\bl}= \exp[-\beta_\text{x}(H_{\bl}- \mathcal{F}^\text{x}_{\bl})]$ in the quasi-static limit. 

Using Eqs.~\eqref{SM:TSwork} and \eqref{SM:fExp} the applied work in the first and the second stroke can now be expressed as 
\begin{subequations}
\begin{align}
    W_1 &= \mathcal{F}^\text{r}_{\bl_1}- \mathcal{F}^\text{r}_{\bl_0} 
    + \be \int_0^{\tau_1} \! dt \; \mathcal{R}^{\mu\nu}_{\bl_t}\dot{\l}^\mu_t \dot{\l}^\nu_t,
    \label{SM:TSwork1}\\
    W_2 &= \mathcal{F}^\text{e}_{\bl_0}- \mathcal{F}^\text{e}_{\bl_1} 
    + \be \int_{\tau_1}^\tau \! dt \; \mathcal{R}^{\mu\nu}_{\bl_t}\dot{\l}^\mu_t \dot{\l}^\nu_t, 
\end{align}
\end{subequations}
where we have used the shorthand notation $\bl_1=\bl_{\tau_1}$. 
With the help of the thermodynamic standard relations $\beta^2\partial_\beta \mathcal{F}_{\bl} = \mathcal{S}_{\bl}$ and $-\beta^3\partial_\beta^2\mathcal{F}_{\bl} = 2\mathcal{S}_{\bl}+\mathcal{C}_{\bl}$, where $\mathcal{S}_{\bl}$ and $\mathcal{C}_{\bl}$ denote the entropy and heat capacity of the system, the free energies at the reservoir temperature in Eq.~\eqref{SM:TSwork1} can be expanded to second order in $A_q$. 
The applied work per cycle, $W=W_1+W_2$, thus becomes 
\begin{align}
    W  =& -\frac{A_q}{\be^2}(\mathcal{S}_{\bl_1}-\mathcal{S}_{\bl_0})
         -\frac{A_q^2}{\be^3}(\mathcal{S}_{\bl_1}-\mathcal{S}_{\bl_0}) 
         -\frac{A_q^2}{2\be^3}(\mathcal{C}_{\bl_1}-\mathcal{C}_{\bl_0})\nonumber\\
       & +  \be \int_0^\tau \! dt \; \mathcal{G}^{\mu\nu}_{\bl_t}\dot{\lambda}^\mu_t\dot{\lambda}^\nu_t, 
       \label{SM:TSwork2}
\end{align}
with $\mathcal{G}^{\mu\nu}_{\bl}=(\mathcal{R}^{\mu\nu}_{\bl} + \mathcal{R}^{\nu\mu}_{\bl})/2$ and $\mathcal{S}_{\bl}$ and $\mathcal{C}_{\bl}$ being taken at the temperature of the environment. 
Comparing Eq.~\eqref{SM:TSwork2} with the adiabatic-response expansion \eqref{MT:FluxExp} of the generalized flux $J_w=W$ yields the kinetic coefficients 
\begin{subequations}\label{SM: Ons coef gen TS}
\begin{align}
    L_{ww} &= \tau \int_0^\tau \! dt \; \mathcal{G}^{\mu\nu}_{\bl_t}\dot{\lambda}^\mu_t\dot{\lambda}^\nu_t\\
    L_{wq} &= -\frac{1}{\be^2}(\mathcal{S}_{\bl_1}-\mathcal{S}_{\bl_0}),\\
    L^q_{wq} &= -\frac{1}{\be^3}(\mathcal{S}_{\bl_1}-\mathcal{S}_{\bl_0}) 
                -\frac{1}{2\be^3}(\mathcal{C}_{\bl_1}-\mathcal{C}_{\bl_0}). 
\end{align}
\end{subequations}

To calculate the heat uptake from the reservoir, we first observe that, in the quasi-static limit, the state of the system changes instantly from $\rho^\text{e}_{\lambda_0}$ to $\rho^\text{r}_{\lambda_0}$ at $t=0$ and from $\rho^\text{r}_{\lambda_1}$ to $\rho^\text{e}_{\lambda_1}$ at $t=\tau_1$. 
These sudden changes occur due to the relaxation dynamics of the system being effectively infinitely fast on the observational time scale. 
They lead to delta-spikes in the heat current $j_t$, which correspond to heat leaks from the system to the reservoir at $t=0$ and from the system to the environment at $t=\tau_1$. 
Hence, the total heat uptake from the reservoir is given by 
\begin{equation}\label{SM:HeatUptake}
    Q = \lim_{\epsilon\rightarrow 0}\int_{0-\epsilon}^{\tau_1-\epsilon}\! dt \; j_t 
      = \lim_{\epsilon\rightarrow 0} (E_{\tau_1-\epsilon} - E_{0-\epsilon}) - W_1,
\end{equation}
where the we have used the first law \eqref{SM:FL}. 
Note that this effect does not play a role for the extracted work, since, assuming that the control protocols $\lambda^\mu_t$ are continuous in $t$, the thermodynamic forces $f^\mu_t$ remain bounded functions of $t$ in the quasi-static limit. 
Hence, no work is done during the switching between reservoir and environment. 

Upon neglecting finite-rate corrections, the heat uptake \eqref{SM:HeatUptake} becomes 
\begin{equation}
    Q= \mathcal{E}^\text{r}_{\bl_1} -\mathcal{E}^\text{e}_{\bl_0}- \mathcal{F}^\text{r}_{\bl_1}+\mathcal{F}^\text{r}_{\bl_0}, 
\end{equation}
where $\mathcal{E}^\text{x}_{\bl}$ denotes the equilibrium internal energy of the system at the inverse temperature $\beta_\text{x}$. 
Expanding this expression to first order in $A_q$ yields 
\begin{equation}\label{SM:TSheatexp}
    Q= \frac{1}{\be}(\mathcal{S}_{\bl_1}-\mathcal{S}_{\lambda_0}) 
     + \frac{A_q}{\be^2}(\mathcal{S}_{\bl_1}-\mathcal{S}_{\bl_0}) + \frac{A_q}{\be^2} \mathcal{C}_{\bl_1} .
\end{equation}
Upon recalling the definition of the heat flux, $J_q= Q/\tau$, and comparing Eq.~\eqref{SM:TSheatexp} with the expansion \eqref{MT:FluxExp}, we can now identify the remaining kinetic coefficients as 
\begin{subequations}\label{SM:KKQq} 
\begin{align}
    L_{qw} &= \frac{1}{\be^2}(\mathcal{S}_{\bl_1}-\mathcal{S}_{\bl_0}), \\
    L^{q}_{qw} &= \frac{1}{\be^3}(\mathcal{S}_{\bl_1}-\mathcal{S}_{\bl_0}) + \frac{\mathcal{C}_{\bl_1}}{\be^3}. 
\end{align}
\end{subequations}
Thus, the geometric figure of merit $\mathcal{Z}$ becomes 
\begin{equation}
    \mathcal{Z} = \frac{L_{qw}^3}{4(L^q_{qw}+L^q_{wq})\mathcal{L}^2} = \frac{(\mathcal{S}_{\bl_1}-\mathcal{S}_{\bl_0})^3}{2 \be^3 (\mathcal{C}_{\bl_1}+\mathcal{C}_{\bl_0})\mathcal{L}^2},
\end{equation}
which is the expression given in Eq.~\eqref{MT:ZTS} of the main text. 

\subsection{Qubit refrigerator}
\subsubsection{Kinetic coefficients}
We consider the qubit-refrigerator described in the main text. 
The system is controlled by two parameters $\lambda^{1}=\lambda$ and $\lambda^{2}\equiv\kappa^{\text{r}}\equiv1-\kappa^{\text{e}}$, which determines the level splitting and the coupling of the qubit to the reservoir and environment, respectively. 
The qubit Hamiltonian is 
\begin{equation}\label{SM: sys H TS}
    H_{\lambda}=\frac{\hbar\Omega}{2}\lambda\sigma_{z},
\end{equation}
where $\sigma_{z}$ is the usual Pauli matrix and $\hbar\Omega$ sets the energy scale. 
The device is driven through a two-stroke cycle as described in the previous section. 
In the first stroke, $\lambda$ is varied linearly from 2 to 1 while $\kappa^{\text{r}}=1$ such that the system is coupled to the reservoir at inverse temperature $\br$. 
In the second stroke, $\lambda$ is linearly returned to its initial value with $\kappa^{\text{r}}=0$, i.e.  the qubit exchanges heat with the environment, whose inverse temperature is $\be$.

For slow driving, the dynamics of the system can be described in terms of the adiabatic master equation \cite{Albash2012},
\begin{equation}\label{SM: Lindblad eq TS}
    \partial_{t}\rho_{t}=-\frac{i}{\hbar}[H_{\lambda_{t}},\rho_{t}]+\mathsf{D}^{\text{r}}_{\boldsymbol{\lambda}_{t}}\rho_{t}+\mathsf{D}^{\text{e}}_{\boldsymbol{\lambda}_{t}}\rho_{t},
\end{equation}
where the dissipation super-operators are
\begin{equation}
    \mathsf{D}^{\text{x}}_{\boldsymbol{\lambda}}\dots=\Gamma\kappa^{\text{x}}\sum_{\alpha=\pm}n^{\alpha}_{\lambda,\beta_{\text{x}}}\big([\sigma_{\alpha}\dots,\sigma^{\dagger}_{\alpha}]+[\sigma_{\alpha},\dots\sigma_{\alpha}^{\dagger}]\big).
\end{equation}
Here, the rate $\Gamma>0$ sets the relaxation time scale and $n^{+}_{\lambda,\beta_{\text{x}}}\equiv1/(e^{\beta_{\text{x}}\hbar\Omega\lambda}-1)$ and $n^{-}_{\lambda,\beta_{\text{x}}}=1+n^{+}_{\lambda,\beta_{\text{x}}}$ are thermal factors; $\sigma_{\pm}=(\sigma_x\pm i\sigma_y)$ are the usual Pauli matrices. 
For fixed control parameters, the unique stationary solution of \eqref{SM: Lindblad eq TS} is given by the Gibbs state $\varrho_{\lambda}^{\text{x}}$.

The equilibrium free energy of the system at reads
\begin{equation}
    \mathcal{F}_{\mathbf{\lambda}}=-\frac{1}{\be}\ln[2\cosh{[\theta \lambda]}],
\end{equation}
where $\theta=\hbar\Omega\beta_{\text{e}}/2$. 
From this result, it is straightforward to calculate the equilibrium entropy and heat capacity,
\begin{subequations}
\begin{align}
    \mathcal{S}_{\lambda}&=\ln\big[2\cosh{[\theta\lambda]}\big]-\theta\lambda\tanh{[\theta\lambda]},\\
    \mathcal{C}_{\lambda}&= \frac{\theta^{2}\lambda^{2}}{\cosh^{2}{[\theta\lambda]}}.
\end{align}
\end{subequations}
Using Eqs.~\eqref{SM: Ons coef gen TS} and \eqref{SM:KKQq}, we thus obtain the kinetic coefficients
\begin{subequations}
\begin{align}
    L_{qw}&= -L_{qw} = \frac{1}{\be^{2}}\Bigl(\ln\big[\cosh{[\theta]}\big]-\ln\big[\cosh{[2\theta]}\big]\Bigr)\nonumber\\
    &\qquad-\frac{\theta}{\be^{2}}\Bigl(\tanh{[2\theta]}-2\tanh{[\theta]}\Bigr),\\
    L_{wq}^{q}&=\frac{\theta^{2}}{2\be^{3}}\left(\frac{4}{\cosh^{2}{[2\theta]}}-\frac{1}{\cosh^{2}{[\theta]}}\right)-\frac{L_{qw}}{\be},\\
    L_{qw}^{q}&=\frac{\theta^{2}}{\be^{3}}\frac{1}{\cosh^{2}{[\theta]}}+\frac{L_{qw}}{\be}. 
\end{align}
\end{subequations}

The coefficient $L_{ww}$ must be found by perturbatively solving the master equation \eqref{SM: Lindblad eq TS}.
To this end, we make ansatz \cite{Cavina2017},
\begin{equation}\label{SM:state ansatz TS}
    \rho_{t}\simeq\varrho_{\lambda}^{\text{e}}+R_{\lambda_{t}}\dot{\lambda}_{t},
\end{equation}
where $R_{\lambda}$ is the first finite-rate correction to the state of the system. 
Substituting \eqref{SM:state ansatz TS} into Eq.~\eqref{SM: Lindblad eq TS} and comparing coefficients gives
\begin{equation}\label{SM:QR}
    R_{\lambda}=\mathsf{L}^{-1}_{\lambda}\partial_{\lambda}\varrho_{\lambda},
\end{equation}
where $\mathsf{L}_{\lambda}\dots=-\frac{i}{\hbar}[H_{\lambda},\dots]+\mathsf{D}_{\boldsymbol{\lambda}}^{\text{r}}\dots+\mathsf{D}_{\boldsymbol{\lambda}}^{\text{e}}\dots$, is the Lindblad generator for the qubit and $\mathsf{L}_{\lambda}^{-1}$ denotes its pseudo-inverse. 
Note that, since we are only interested in the coefficient $L_{ww}$, which describes the isothermal work contribution, we can set $\br=\be$.
The generator $\mathsf{L}_\lambda$ does therefore not depend on the coupling parameter $\kappa$ here. 

Using the result \eqref{SM:QR}, we can evaluate the expression \eqref{SM:Deffj} for the generalized force $f_t$, which gives the adiabatic-response coefficient 
\begin{equation}
    \mathcal{R}^{11}_{\lambda}=\mathcal{R}_{\lambda}=\frac{\hbar\Omega\theta}{4\be\Gamma}\frac{\tanh{[\theta\lambda_{t}]}}{\cosh^{2}{[\theta\lambda_{t}]}}. 
\end{equation}
We thus have 
\begin{equation}
    L_{ww}=\frac{\hbar\Omega\theta\tau}{4\be\Gamma}\int_{0}^{\tau}dt\,\frac{\tanh{[\theta\lambda_{t}]}}{\cosh^{2}{[\theta\lambda_{t}]}}\geq\mathcal{L}^{2}
\end{equation}
and the thermodynamic length is given by 
\begin{equation}
    \mathcal{L}=\frac{\hbar\Omega\theta\tau}{4\be\Gamma}\oint_{\gamma}\sqrt{\tanh{[\theta\lambda]}/\cosh^{2}{[\theta\lambda]}}\, d\lambda
\end{equation}
accordint to Eq.~\eqref{MT:TL}. 
The optimal speed function $\phi_t$ can now be found by numerically solving Eq.~\eqref{MT:SF} with $\mathcal{G}^{11}_\lambda = \mathcal{R}_\lambda$. 

\subsubsection{Numerical Calculations}
To obtain the exact dynamics of the qubit-refrigerator, we make the ansatz
\begin{equation}
    \rho_{t}=\mathbb{I}/2 + \rho^{z}_{t}\sigma_{z}
\end{equation}
for the state of the system, where $\mathbb{I}$ denotes the identity matrix. 
Substituting this ansatz into Eq.~\eqref{SM: Lindblad eq TS} yields differential equation
\begin{equation}\label{SM:TS dynamics}
    \partial_{t}\rho_{t}^{z}=-\Gamma\sum_{x=\text{e},\text{r}}\kappa^{\text{x}}_{t}\big[2k^{+}_{\lambda_{t},\beta_{\text{x}}}\rho^{z}_{t}-k^{-}_{\lambda_{t},\beta_{\text{x}}}\big],
\end{equation}
with $k^{\pm}_{\lambda_{t},\beta_{\text{x}}}=(n_{\lambda_{t},\beta_{\text{x}}}^{+}\pm n_{\lambda_{t},\beta_{\text{x}}})$. 
After specifying the control protocols $\boldsymbol{\lambda}_t$ as described before, we solve this equation numerically for $\rho_{t}^{z}$ with periodic boundary conditions. 
Following Eq.~\eqref{MT:WQ}, the applied work and heat uptake can then be found as 
\begin{subequations}
\begin{align}
    W&=\hbar\Omega\int_{0}^{\tau}dt\, \rho_{t}^{z}\dot{\lambda}_{t},\\
    J_{q}&=-\frac{2\hbar\Omega\Gamma}{\tau}\int_{0}^{\tau_{1}} dt\,\frac{\rho_{t}^{z}\lambda_{t}}{\tanh{[\theta_{\text{r}}\lambda_{t}]}},
\end{align}
\end{subequations}
where $\theta_{\text{r}}=\hbar\Omega\br/2$. 
The cooling power and COP of the qubit-refrigerator are obtained by substituting the solution of \eqref{SM:TS dynamics} into the expressions above for a range of cycle times $\tau$.

\newcommand{\x}{\text{x}}
\renewcommand{\r}{\text{r}}
\newcommand{\y}{\text{y}}
\newcommand{\e}{\text{e}}
\newcommand{\ket}[1]{|#1\rangle}
\newcommand{\bra}[1]{\langle #1 |}
\newcommand{\braket}[2]{\langle #1 | #2\rangle}

\section{Coherent transport}
In this section, we consider coherent conductors that are subject to slow periodic driving and coupled to two thermal baths with a small temperature difference, to which we refer as reservoir and environment. 
We first outline how the kinetic coefficients $L_{xy}$ and $L_{xy}^q$ can be obtained for such systems in general. 
In the second part, we focus on the Aharonov-Bohm refrigerator discussed at the end of the main text. 

\subsection{Kinetic Coefficients}
We first recall the expressions for the generalized forces and the heat current from the reservoir into the system, 
\begin{subequations}\label{SM:fjScatt}
\begin{align}
    f^\mu_t &= - \int_0^\infty \!\!\! dE \sum_{\x=\r,\e}\bra{\psi_{E,t}^{\x}}\partial_\mu H_{\bl_t}\ket{\psi_{E,t}^{\x}} g^\text{x}_E\quad\text{and}\\
    j_t &= \int_0^\infty \!\!\! dE\sum_{\x=\r,\e} \bra{\psi_{E,t}^{\x}}J_{\bl_t} \ket{\psi_{E,t}^{\x}} g^\x_E. 
\end{align}
\end{subequations}
Here, $g^\text{x}_E\equiv 1/[1+e^{\beta_\text{x}(E-\mu)}]$ denotes the Fermi function of either the reservoir ($\x=\r$) or the environment ($\x=\e$), where $\mu$ is the chemical potential. 
Furthermore, 
\begin{subequations}
\begin{align}
    H_{\bl} &= \frac{p^2}{2m} + V_{\bl}(x),\\
    J_{\bl} &= -\frac{1}{4,}\{(H_{\bl}-\mu),\{p,\delta[x-r_\r]\}\}
\end{align}
\end{subequations}
are the single-carrier Hamiltonian and the heat current operator, where $x$ and $p$ are the usual position and momentum operators, $m$ is the carrier mass and $V_{\bl}$ is an externally tunable scattering potential, which vanishes outside a bounded scattering region. 
Curly brackets denote anti-commutators.  
The Floquet scattering states $\ket{\psi_{E,t}^\x}$ describe a carrier with energy $E$ that enters the conductor from the reservoir or the environment \cite{Moskalets2002,Brandner2020b}. 
These states are periodically dependent on $t$ and satisfy the Floquet-Schr\"odinger equation 
\begin{equation}\label{SM:FSE}
    (H_{\bl_t}-i\hbar \partial_t)\ket{\psi^\x_{E,t}}= E\ket{\psi^\x_{E,t}}
\end{equation}
with respect to the boundary conditions
\begin{equation}\label{SM:FSSBC}
    \braket{r_\y}{\psi^\x_{E,t}}= \delta_{\x\y}w^-_E[r_\y]+ S^{\y\x}_{E,t}w^+_E[r_\y]. 
\end{equation}
Here, $w_\pm[r]= \xi_E e^{\pm ik_E r}$ are incoming $(-)$ or outgoing $(+)$ plane waves with wave number $k_E=[2mE/\hbar^2]^\frac{1}{2}$ and $\xi_E=[m/2\pi k_E \hbar^2]^\frac{1}{2}$ is a normalization factor. 
The spatial coordinate $r_\x\geq 0$ parameterizes the lead that connects the scattering region with the bath $\x$. 
For simplicity, we assume that the leads support only a single transport channel. 
Hence, they are effectively one-dimensional. 
Finally, the Floquet scattering amplitude $S^{\y\x}_{E,t}$, which is periodic in $t$, corresponds to the probability amplitude that a carrier with energy $E$ is transmitted from the lead $\x$ to the lead $\y$ at the time $t$. 

In the quasi-static limit, where the driving is infinitely slow compared to the dwell time of carriers inside the scattering region, the Floquet scattering states $\ket{\psi^\x_{E,t}}$ reduce to the frozen scattering states $\ket{\phi^\x_{E,\bl}}$, which satisfy 
\begin{subequations}
\begin{align}
    & H_{\bl} \ket{\phi^\x_{E,\bl}} = E\ket{\phi^\x_{E,\bl}} \quad\text{and}\quad\\
    & \braket{r_\y}{\phi^\x_{E,\bl}}= \delta_{\x\y}w^-_E[r_\y]+ \mathcal{S}^{\y\x}_{E,\bl}w^+_E[r_\y]
\end{align}
\end{subequations}
with $\mathcal{S}^{\y\x}_{E,\bl}$ being the frozen scattering amplitudes. 
Finite-rate corrections to the Floquet scattering states can now be calculate by means of adiabatic perturbation theory \cite{Thomas2012,Bode2012}. 
In first order, this approach yields
\begin{equation}\label{SM:FSfirstorder}
    \ket{\psi^\x_{E,t}}= \ket{\phi^\x_{E,\bl_t}} - i\hbar (G^+_{E,\bl_t})^2(\partial_\mu H_{\bl_t})\ket{\phi^\x_{E,\bl_t}}\dot{\l}_t^\mu
\end{equation}
with the retarded Greens function 
\begin{equation}
    G^+_{E,\bl} = \frac{1}{E-H_{\bl}+i\epsilon}. 
\end{equation}

\newcommand{\dS}{\dot{\mathcal{S}}}
\newcommand{\av}[1]{\bigl\llangle #1\bigr\rrangle}

Upon inserting the approximation \eqref{SM:FSfirstorder} into the Eqs.~\eqref{SM:fjScatt}, the generalized forces $f^\mu_t$ and the heat current $j_t$ can be calculated to first order in the rates $\dot{\l}^\mu_t$. 
The fluxes $J_x$ can then be obtained from Eqs.~\eqref{MT:FluxAff} and \eqref{MT:WQ}. 
Expanding in the thermal gradient $A_q$ and comparing the result with the adiabatic-response relations  
\begin{align}
    J_w & = W=  L_{ww}A_w + L_{wq}A_q + L^q_{wq}A_q^2,\\
    J_q & = Q/\tau = L_{qw}A_w + L_{qq}A_q + L^q_{qw}A_w A_q,
\end{align}
finally gives the kinetic coefficients $L_{xy}$ and $L^q_{xy}$. 
This procedure can be carried out using standard manipulations of scattering theory. 
The final results are given by 
\begin{subequations}\label{SM:ScatKC}
\begin{align}
    L_{ww} &= -\frac{\tau \hbar}{2} \int_0^\infty \!\!\! dE \;  
    \frac{g'_E}{E-\mu} \sum_{\x,\y, = \e,\r}\av{\bigl|\dS^{\x\y}_{E,\bl}\bigr|^2},\\
    L_{qq} &=-\frac{1}{\tau \hbar} \int_0^\infty \!\!\! dE \;
    g'_E \; \av{\bigl|\mathcal{S}^{\r\e}_{E,\bl}\bigr|^2},\\
    L_{wq} &=\int_0^\infty \!\!\! dE \; g'_E \sum_{\x = \e,\r} 
    \text{Im}\bigl[\av{\dS^{\x\r}_{E,\bl}\mathcal{S}^{\x\r\ast}_{E,\bl}}\bigr],\\
    L_{qw} &= - \int_0^\infty \!\!\! dE \; g'_E \sum_{\x = \e,\r} 
    \text{Im}\bigl[\av{\dS^{\r\x}_{E,\bl}\mathcal{S}^{\r\x\ast}_{E,\bl}}\bigr],\\
    L^q_{wq} &= -\frac{1}{2} \int_0^\infty \!\!\! dE \; g''_E \sum_{\x = \e,\r} 
    \text{Im}\bigl[\av{\dS^{\x\r}_{E,\bl}\mathcal{S}^{\x\r\ast}_{E,\bl}}\bigr],\\
    L_{qw}^q &= \frac{1}{\be} \int_0^\infty \!\!\! dE \;
    \bigl(g'_E+\be g''_E\bigr)\;\text{Im}\bigl[\av{\dS^{\r\r}_{E,\bl}\mathcal{S}^{\r\r\ast}_{E,\bl}}\bigr], 
\end{align}
\end{subequations}
where $\llangle\cdots\rrangle=\frac{1}{2\pi}\int_0^\tau dt \; \cdots$ and we have introduced the abbreviations 
\begin{align}
    g'_E & = - g^\e_E(1-g^\e_E)(E-\mu),\\
    g''_E & =  g^\e_E(1-g^\e_E)(1-2g^\e_E)(E-\mu)^2. 
\end{align}
Dots indicate total time derivatives. 
Notably, the expressions \eqref{SM:ScatKC} depend only on the frozen scattering amplitudes, which, 
unless time-reversal symmetry is broken, obey $\mathcal{S}^{\x\y}_{E,\bl}=\mathcal{S}^{\y\x}_{E,\bl}$. 
Hence, the coefficients $L_{wq}$ and $L_{qw}$ satisfy the Onsager symmetry \eqref{MT:OS}. 
For details on the derivation of the Eqs.~\eqref{SM:ScatKC}, we refer the reader to the existing literature \cite{Moskalets2002,Bode2012,Thomas2012,Potanina2021,Brandner2020b}. 

\subsection{Aharonov-Bohm Refrigerator}
\subsubsection{Kinetic Coefficients}
We now move on to the mesoscopic refrigerator shown in Fig.~2 of the main text. 
This model was originally introduced in Ref.~\cite{Avron2001} as an example for an ideal quantum pump. 
Its frozen scattering amplitudes are given by 
\begin{equation}
    \mathcal{S}^{\r\e}_{E,\bl}=\mathcal{S}^{\e\r}_{E,\bl} =0, \;\;
    \mathcal{S}^{\r\r}_{E,\bl} = \l^1 + i\l^2, \;\;
    \mathcal{S}^{\e\e}_{E,\bl} = \l^1 - i\l^2,
\end{equation}
where the two control parameters $\l^1$ and $\l^2$ correspond to the real and the imaginary part of the Aharonov-Bohm phase that is picked up by a carrier passing through the ring, i.e. $e^{iq\Phi/\hbar c}=\l^1+i\l^2$.
A linearly increasing magnetic flux, $\Phi_t= \hbar c \omega t/q$, thus leads to the driving protocols $\l^1_t = \cos[\omega t]$ and $\l^2_t = \sin[\omega t]$. 
For these protocols, it is straightforward to calculate the kinetic coefficiens \eqref{SM:ScatKC}. 
Specfically, we find 
\begin{subequations}\label{SM: Ons coef AB}
\begin{flalign}
    L_{ww}&=\frac{2\pi \hbar \varphi}{\be(1+\varphi)},
    \\
    L_{qw}&=-L_{wq}=-\frac{\mu \varphi}{\be(1+ \varphi)}+\frac{1}{\be^{2}}\log{[1+\varphi]},
    \\
    L_{wq}^{q}&=\frac{\mu \varphi(2+2\varphi-\be\mu)}{2\be^2(1+\varphi)^{2}}-\frac{1}{\be^{3}}\log{[1+\varphi]},
    \\
    L_{qw}^{q}&=-\frac{\mu \varphi(1+\varphi-\be\mu)}{\be^2(1+\varphi)^{2}}+\frac{1}{\be^{3}}\log{[1+\varphi]}
\end{flalign}
\end{subequations}
with $\varphi= e^{\be\mu}$. 
Note that the coefficient $L_{qq}$ is zero, since the quasi-static heat current vanishes as discussed in the main text.  
Furthermore, $L_{ww}$ equals the square of the thermodynamic length $\mathcal{L}$, since the linear parameterization $\phi_t =t$ of the control path is optimal for this model. 

\subsubsection{Numerical Calculations}

\begin{figure}
\includegraphics[scale=0.7]{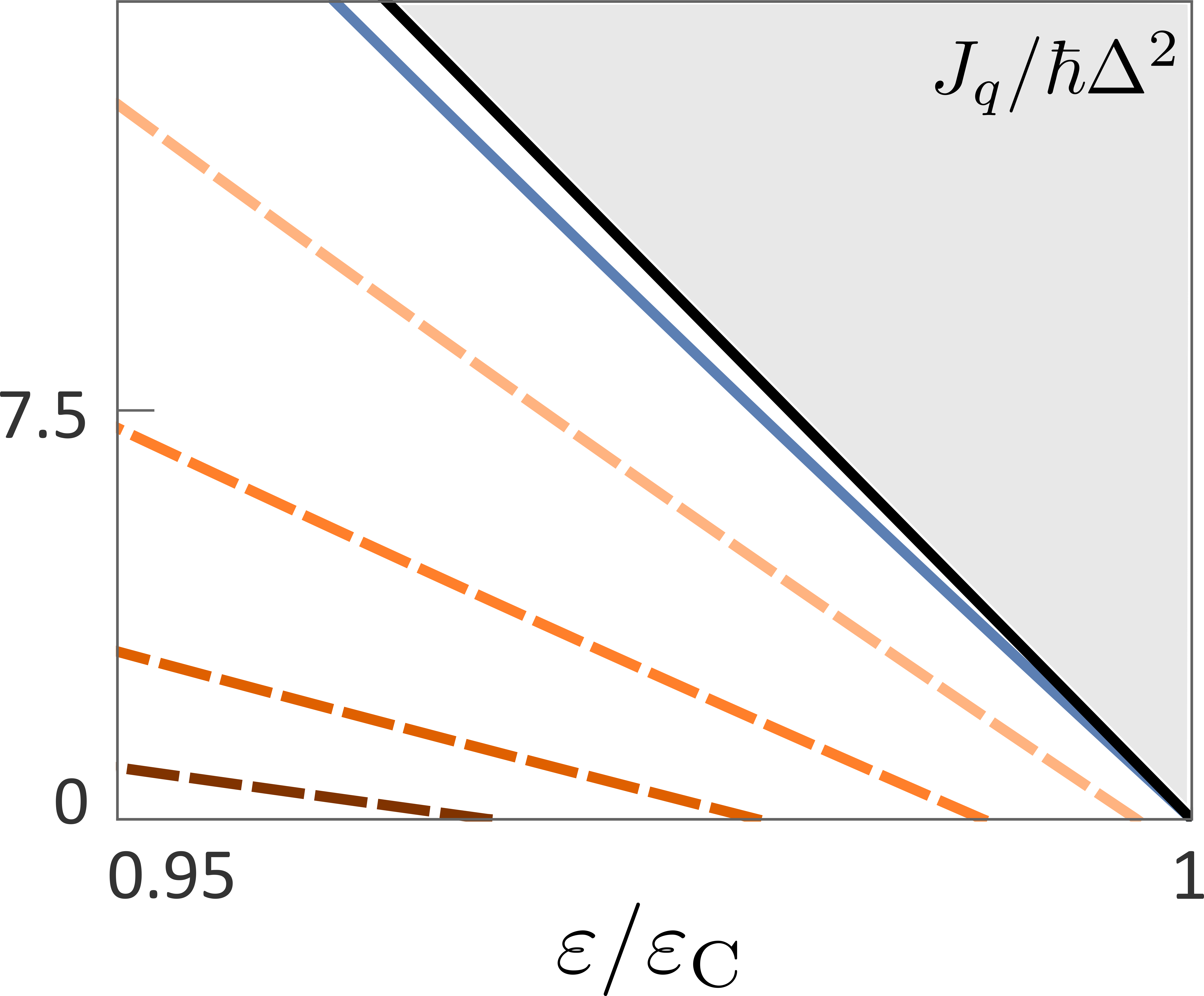}
\caption{Aharonov-Bohm refrigerator revisited. 
The cooling power in units $10^{-4}\hbar\Delta^{2}$ is plotted as a function of the normalised COP.
For $A_w\rightarrow 0$, the linear bound \eqref{SM: AB lin}, shown by grey area, is approached asymptotically by the blue line, which corresponds to $\mu=0$.
By contrast, the dashed lines, for which we have set $\mu=0.5\hbar\Delta,...,2\hbar\Delta$ from top to bottom, do not reach the Carnot limit. 
For all curves, we have fixed the thermal gradient as $A_{q}=-\be/24$ and scaled the driving frequency from $\omega = 10^{-6} \Delta$ to $\omega = 5\times10^{-3}\Delta$. 
\label{Fig:TLR}}
\end{figure}

The exact Floquet scattering amplitudes for the Aharonov-Bohm device were calculated in App.~D. of Ref.~\cite{Potanina2021}.
Using these results, the generalized fluxes can be written as 
\begin{subequations}\label{SM:ABFlux}
\begin{align}
     J_q &= \frac{1}{2\pi\hbar}\int_0^\infty \!\!\! dE \; (E-\mu)\bigl((g^\r_{E}-g^\r_{E+\hbar\omega})\nonumber\\
    &\hspace{3cm} +  (g^\r_{E+\hbar\omega}-g^\e_E)T_E\bigr),\\
    J_w &= -\int_0^\infty \!\!\! dE \; (g^\r_{E+\hbar\omega}-g^\e_E)
\end{align}
\end{subequations}
with the transmission and reflection coefficients 
\begin{equation}
    T_E = \frac{|\hat{\mathbf{1}}^t\mathbb{A}_{E+\hbar\omega}\mathbb{A}_{E}^{-1}\mathbf{1}|^2}{
    |\mathbf{1}^t\mathbb{A}_{E+\hbar\omega}\mathbb{A}_{E}^{-1}\mathbf{1}|^2}, \quad
    R_E = \frac{4}{|\mathbf{1}^t\mathbb{A}_{E+\hbar\omega}\mathbb{A}_{E}^{-1}\mathbf{1}|^2}.
\end{equation}
Here, $\mathbf{1}=(1,1)^t$ and $\hat{\mathbf{1}^t=(1,-1)^t}$ are vectors and the matrix $\mathbb{A}_E$ is defined as 
\begin{equation}
\mathbb{A}_E
    =\frac{1}{\xi_E}\left[ 
    \begin{array}{ll}
     \text{Ai}_E    &  \text{Bi}_E\\
     -i\text{Ai}'_E   & -i \text{Bi}'_E
    \end{array}
    \right]
\end{equation}
with 
\begin{subequations}
\begin{align}
    \text{Xi}_E & = \text{Xi}[(\omega/\Delta)^{-\frac{2}{3}}(-E/\hbar\Delta)],\\
    \text{Xi}'_E &= (\omega/\Delta)^\frac{1}{3}(\hbar\Delta/E)^\frac{1}{2}
    \text{Xi}'[(\omega/\Delta)^{-\frac{2}{3}}(-E/\hbar\Delta)]. 
\end{align}
\end{subequations}
In these expressions, $\text{Xi}=\text{Ai},\text{Bi}$ and $\text{Xi}'=\text{Ai}',\text{Bi}'$ denote the standard Airy functions and their derivatives. 
The parameter $2\pi/\Delta = 4\pi m l^2/\hbar^2$ corresponds to the typical dwell time of carriers inside the device.  
Hence, the adiabatic-response regime is defined by the condition $\omega/\Delta \ll 1$.
We recall that $\xi_E$ was defined below Eq.~\eqref{SM:FSSBC} and that $l$ is the circumference of the Aharonov-Bohm ring, see Fig.~2 of the main text. 
Upon evaluating the Eqs.~\eqref{SM:ABFlux} numerically, we obtain the cooling power and efficiency of the Aharonov-Bohm refrigerator for arbitrary driving frequencies. 
These results provide us with a benchmark for the adiabatic-response theory.

\subsubsection{Linear Trade-off Relations}

The formulas \eqref{SM:ScatKC} for the kinetic coefficients of coherent conductors confirm that there is in general no obvious relation between the second-order coefficients $L^q_{wq}$ and $L^q_{qw}$. 
However, there exist special situations, where $L^q_{wq}=-L^q_{qw}$. 
Under this condition, the first term in the expansion of the normalized COP, see Eq.~\eqref{MT:COPExp}, vanishes and we are left with the trivial relation 
\begin{equation}
    \varepsilon/\varepsilon_{\text{C}}=1+\frac{L_{ww}A_{w}}{L_{qw}A_{q}}. 
\end{equation}
This result indicates that the Carnot bound is attained in the quasi-static limit $A_w\rightarrow 0$ irrespective of $A_{q}$. 
Furthermore, since the cooling power is $J_{q}=L_{qw}A_{w}$ in leading order, we obtain the linear power-COP trade-off relation 
\begin{equation}\label{SM: AB lin}
    J_{q}\leq Y(\varepsilon_{\text{C}}-\varepsilon)/\varepsilon_{\text{C}},
\end{equation}
where $Y=L_{qw}^{2}|A_{q}|/L_{ww}\geq 0$ \footnote{It is straightforward to derive the analogous trade-off relation for heat-engines, $P\leq Y(\eta_{\text{C}}-\eta)$}. 

We now return to the Aharonov-Bohm refrigerator.
The sum of the second-order coefficients is
\begin{equation}
    L_{wq}^{q}+L_{qw}^{q}=\frac{\mu^{2} e^{\be\mu}}{2\be(1+e^{\be\mu})^{2}}. 
\end{equation}
This quantity vanishes for $\mu\rightarrow 0$. 
Hence, in this limit, we expect that the device approaches its Carnot bound for any sufficiently small thermal gradient $A_q$ as $A_w$ goes to zero. 
Furthermore, its power output should decay linearly towards the Carnot limit. 
This behaviour is indeed confirmed by our numerical calculations, see Fig.~\ref{Fig:TLR}.